\documentclass[aps,amssymb,amsmath, twocolumn, pra, superscriptaddress]{revtex4}
\usepackage{graphicx}
\usepackage{epsfig}
\usepackage{dcolumn}
\usepackage[up]{subfigure}
\usepackage{float}
\usepackage{dsfont}	
\usepackage{relsize}

\usepackage{epstopdf}

\usepackage{float}
\usepackage{ragged2e}
\usepackage{braket}
\usepackage[font=small,labelfont=bf,justification=raggedright, format=plain]{caption}
\usepackage{setspace}
\usepackage[colorlinks, citecolor = magenta]{hyperref}
\makeatletter

\begin{document}


\title{Harnessing Brillouin Interaction in Rare-earth Aluminosilicate Glass Microwires for Optoelectromechanic Quantum Transduction}

\author{Mrittunjoy Guha Majumdar}
\email{mguhamajumdar@fas.harvard.edu}
\affiliation{Quantum Optics \& Quantum Information, Department of Instrumentation and Applied Physics, Indian Institute of Science, Bengaluru 560012, India}
\author{C. M. Chandrashekar}
\email{chandracm@iisc.ac.in}
\affiliation{Quantum Optics \& Quantum Information, Department of Instrumentation and Applied Physics, Indian Institute of Science, Bengaluru 560012, India}
\affiliation{The Institute of Mathematical Sciences, C. I. T. Campus, Taramani, Chennai 600113, India}
\affiliation{Homi Bhabha National Institute, Training School Complex, Anushakti Nagar, Mumbai 400094, India}

\begin{abstract}
Quantum transduction, the process of converting quantum signals from one form of energy to another is a key step in harnessing different physical platforms and the associated qubits for quantum information processing. Optoelectromechanics has been one of the effective approaches to undertake transduction from optical-to-microwave signals, among others such as those using atomic ensembles, collective magnetostatic spin excitations, piezoelectricity and electro-optomechanical resonator using Silicon nitride membrane. One of the key areas of loss of photon conversion rate in optoelectromechanical method using Silicon nitride nanomembranes has been those in the electro-optic conversion. To address this,  we propose the use of Brillouin interactions in a fiber mode that is allowed to be passed through a fiber taper in rare-earth Aluminium glass microwires. It suggests that we can efficiently convert a $195.57$ THz optical signal to a $325.08$ MHz microwave signal with the help of Brillouin interactions, with a whispering stimulated Brillouin scattering mode yielding a conversion efficiency of $\sim45$\%.
\end{abstract}



\maketitle

\section{Introduction}

Integrated quantum information processing with localised quantum computing systems and distributed quantum communication architecture requires the ability to transduce efficiently between the relevant platforms\,\cite{lauk2020perspectives, rakher2010quantum, vandevender2007quantum, zhang2018quantum}. Functionally, that often means a parametric conversion between distinct regimes in which the systems are operating. For instance, if we utilise superconducting qubits for quantum computing and undertake quantum communication using light, we would need optical-to-microwave quantum transduction capability\, \cite{witmer2020silicon, forsch2020microwave}. Such transduction usually involves the faithful transfer of quantum information encoded in a set of bosonic operators ($\{\hat{a}_{j}^{\dagger}\}$) to another set of bosonic operators ($\{\hat{b}_{j}^{\dagger}\}$). This could be possible with distinct bosons, such as photons and phonons, or the same kind of boson but with dissimilarity in the values within atleast one degree-of-freedom, such as electromagnetic field modes at non-identical frequencies\,\cite{lau2019high, imade2021gigahertz, jiang2020efficient}. 

When it comes to optical-to-microwave transduction, the primary problem is the large difference between the mode-frequencies that lead to off-resonant interactions\,\cite{peairs2020continuous, lauk2020perspectives}. Intermediate systems, such as atomic ensembles\,\cite{imamouglu2009cavity, hafezi2012atomic, verdu2009strong, o2014interfacing, fernandez2015coherent, williamson2014magneto} and magnons\,\cite{everts2019microwave, hisatomi2016bidirectional} that couple to both modes - optical and microwave signals, can facilitate the bridging of the energy-gap\,\cite{holzgrafe2020cavity}. Optoelectromechanical approaches can provide an efficient interface for undertaking quantum transduction, with optomechanics using elements such as photoelasticity\,\cite{aben1993integrated, mueller1938theory, aspelmeyer2014cavity, bowen2015quantum} and electromechanics using an interfacing capacitive element\,\cite{teufel2011circuit, haghighi2018sensitivity, jones2013electromechanics, li2020stationary, wu2020microwave}. Double-cavity optomechanical systems are the centrepoint of a standard method for realisation of optical-to-microwave transduction. Such a system couples a microwave resonator mode $C_{1}$ with resonance frequency $\omega_{C,1}$ to the vibrational mode of a mechanical resonator with frequency $\omega_{m}$, that is, simultaneously coupled to an optical cavity mode $C_{2}$ with resonance frequency $\omega_{C,2}$. 

The mediating phononic mode in the optoelectromechanical realisation of quantum transduction is usually generated in a Silicon nitride  (SiN) nanomembrane. It has been experimentally demonstrated by Andrews et al.\,\cite{andrews2014bidirectional} wherein SiN nanomembrane is used as one of the mirrors in a Fabry-Perot cavity while constituting the capacitive coupling to a superconductor microwave resonator.
In the experiment, bidirectional transduction was performed with 10\% net efficiency, albeit with about 1700 noise quanta generated at cryogenic temperatures. The efficiency of conversion was limited by microwave resonator losses and imperfect mode-matching, although the two main sources of noise were the thermal vibrational noise and spurious mechanical modes in the membrane. In another experiment,  mechanically mediated electro-optic conversion in an alternative feed- forward framework  was used to demonstrate  microwave to optical conversion with 47\% conversion efficiency\,\cite{HBU2018}. 

For efficient transduction, we must remove these spurious mechanical modes, have low-loss optical transmission platform as well as large opto- and electromechanical coupling rates.  The use of photonic fibers optimises transmission of optical quantum signals, and thereby the use of the same in any transduction framework can help achieve high efficiency in the process. In this paper, we propose the use of traveling-wave optomechanical interactions known as Brillouin interactions using rare-earth aluminosilicate glass microwires. A pressure or force density on the material results from electrostriction or radiation pressure when optical modes of a waveguide interfere with one another. In the event that there is conservation of momentum and energy during the interaction, this pressure then stimulates an acoustic wave. As a result of the acoustic field's stresses altering the medium's dielectric characteristics and perturbing its boundaries, light is essentially scattered across distinct optical modes. The new proposed scheme using rare-earth aluminosilicate glass microwires also offers a conversion rate of $\approx 45\%$, far higher than any other scheme using silicon based membrane as transduction medium.

\section{Physical Interactions at Coupling Interfaces} 

Our scheme for undertaking optical-to-microwave transduction involves two distinct physical interactions at the coupling interfaces: opto-acoustic interactions using Brillouin scattering and dielectric-modulated magnetic response at the acousto-electric interface \cite{damzen2003stimulated, sipe2016hamiltonian, sturmberg2019finite}. 
 
\subsection{Opto-acoustic Interaction with Brillouin Scattering}

\begin{figure}
    \centering
    \includegraphics[scale=0.6]{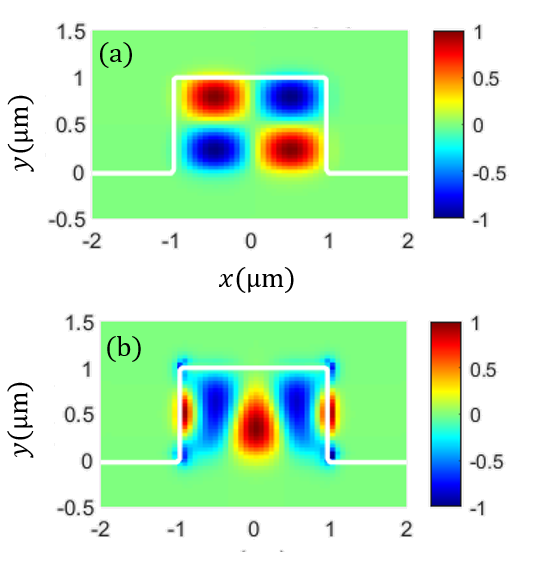}
    \caption{Illustration of effective optical mode in a 2D waveguide for effective refractive index of microwire, $n_{\mathrm{eff}} = 2.841$, with (a) being $\mathrm{Re}(E_x)$ and (b) being $\mathrm{Re}(E_y)$.}
    \label{fig: waveguide}
\end{figure}

Certain underlying processes in optically isotropic material substrates result in opto-acoustic interactions \cite{sturmberg2019finite}. The electromagnetic field can produce mechanical strains in the waveguide through electrostriction \cite{sundar1992electrostriction}. The converse, in the phenomenon of photoelasticity, is observed when stresses cause localized variations in the dielectric permittivity \cite{mueller1938theory}. Waveguide boundaries may physically shift as a result of radiation pressure caused by the electromagnetic waves reflecting off of the boundaries of the structure, which in turn drives acoustic waves. A localised variation in the electromagnetic characteristics is caused by the mobility of structural boundaries brought about by mechanical vibrations. Stimulated Brillouin Scattering (SBS) is the nonlinear phenomenon whereby an injected pump wave is scattered by an acoustic vibration into the Stoke and anti-Stoke wave components\,\cite{damzen2003stimulated, pant2011chip, garmire2017perspectives, bai2018stimulated}. In 1972, Ippen et al. achieved low-threshold stimulated Brillouin scattering with input powers as low as 1 W\,\cite{ippen1972stimulated}. 
\\
\\
When we couple and guide a coherent laser beam into an optical microwire, the light gives rise to as well as experiences various kinds of elastic waves that have similar frequencies. While light is sensitive only to longitudinal and shear bulk acoustic waves in standard optical fibers, the light and the evanescent field access the outer-surface as well, in the case of a sub-wavelength optical fiber, thereby causing the shaking of the wire due to electrostriction which leads to surface acoustic waves (SAWs) being generated. The effective refractive index along the microwire varies periodically due to these ripples, leading to Bragg scattering of the light in the backward direction as per phasematching condition. The velocity of the bulk and surface acoustic waves differ significantly, with the surface wave travelling at a velocity of around 0.9 that of a shear wave and giving rise to other optical sidebands. Beugnot et al. found the surface acoustic wave mode to travel at a velocity of 3400 m/s and a frequency of 6 GHz, having studied the generation of SAWs in an 8 cm. long silica optical microwire that was drawn from a single-mode fiber using the heat-brush method\,\cite{beugnot2014stimulated}. 
\\
\\
Photon silica microwires, fabricated by tapering optical fibers, have been seen to support stimulated Brillouin scattering\,\cite{beugnot2014stimulated}. It is found that rare-earth materials like Lanthana and Ytterbia have a finite effect on the Brillouin characteristics of silica-based oxide glass optical fibers, with emergent attributes such as a wide Brillouin spectral width, low acoustic velocity and a negative photoelastic constant\,\cite{dragic2014brillouin}. Brillouin processes can alternatively be visualised as the formation and destruction of quasi-particles, in this case photons and phonons\,\cite{eggleton2019brillouin}. It is typically advantageous to differentiate interactions based on whether there is an absorption or emission of the phonon. A phonon is produced when a high-frequency photon undergoes the Stokes Brillouin transition and changes into a lower-frequency Stokes photon\,\cite{tang1966saturation, hill1976cw}. This transition may happen naturally or may be stimulated. Anti-Stokes transitions, wherein a phonon is absorbed while a lower-frequency photon is changed to a higher-frequency photon, are also conceivable\,\cite{enzian2022brillouin}. Brillouin processes can be triggered by either an optical seed from an auxiliary optical field at the Stokes frequency or by thermal phonons in the waveguide, which scatter the pump photons\,\cite{bergman2018brillouin}.
\\
\\
In almost every system of practical importance, it is infeasible to compute Brillouin scattering from first principles, for example by considering Maxwell's equations nonlinearly linked to the Christoffel equation. This is caused by a division of scales. To start, the optical and acoustic problems' temporal scales often differ by five orders of magnitude. Second, interaction lengths on the millimetre scale are necessary for integrated waveguides that have nonpareil and superlative gain coefficients to give a noticeable overall response - the system scale is four orders of magnitude larger than the acoustic wave length. A coupled mode description is especially appropriate to Brillouin problems due to this scaling divergence\, \cite{behunin2018fundamental}. We can express the acoustic and optical fields as eigenmodes of the waveguide, which are weighted by envelop functions - $a_n(z,t)$ for each optical mode and $b(z,t)$ for the acoustic field, thereby giving us the states\,\cite{wolff2015stimulated},
\begin{multline}
\vert\psi(x,y,z,t)\rangle = \sum_{n}a_{n}(z,t)\vert\Psi_{n}(x,y)\rangle e^{i k_n z-i\omega_n t}+c.c,\\ \vert\phi(x,y,z,t)\rangle = b(z,t)\vert\Phi(x,y)\rangle e^{i qz-i\Omega_n t}+c.c,
\end{multline}
where $c.c$ represents complex conjugation. The index $n$ are to characterise the distinct optical modes, $q$ and $k_n$ are the acoustic and optical wavenumbers, $\Omega$ and $\omega_n$ are the acoustic and optical frequencies.  Usually, the scope of applying such modal expansion depends on whether the amplitudes (envelop functions) have a slower variation than their respective carrier propagation terms. The basis functions $\vert\Psi_{n}(x,y)\rangle$ are bound solutions to the eigenproblem $(\nabla_\perp + i\beta\hat{z})\times (\nabla_\perp + i\beta\hat{z}) \times \vert\Psi_{n}(x,y)\rangle = \varepsilon \mu_0 \omega^2 \vert\Psi_{n}(x,y)\rangle$ while $\vert\Phi(x,y)\rangle$ satisfy $\rho\Omega^2\vert\Phi\rangle_i +\sum_{jkl}(\nabla_\perp + iq\hat{z})_j c_{ijkl}(\nabla_\perp + iq\hat{z})_k\vert\Phi\rangle_l = 0.$ The first eigen-formulation arises from the manner in which optical fields evolve by the electromagnetic wave equations in terms of the electric field ($\textbf{E}$) and electric induction field ($\textbf{D}$): $\nabla\times\nabla\times \textbf{E}= -\mu_0\partial_t^2 \textbf{D}$, $\textbf{D} = \varepsilon E$ and $\varepsilon = \varepsilon_r \varepsilon_0$, with $\varepsilon$ being the dielectric constant. 
In Fig.\,\ref{fig: waveguide} effective optical mode in a 2D waveguide is show when the effective refractive index $n_{\mathrm{eff}} = 2.841$ of a rectangular waveguide. The second equation arises from the acoustic wave equation for a mechanical displacement field\,\cite{auld1973acoustic}.
\\
\\
The optical field satisfies the Maxwell's equations: $\nabla\times E = -\delta_t B$, $\nabla\times H = -\delta_t D + J$, where $B$ and $D$ are the magnetic and electric induction fields, while $H$ and $E$ are the magnetic and electric fields, respectively. $J$ denotes the dissipative current that arises because of Ohmic losses. The optical side of the composite system is characterised by, $(\delta_t +v_n\delta_z+\Gamma_n)a_n=\frac{1}{E_n}i\sum_{n'}e^{i(\phi_n-\phi_{n'})}[e^{i(qz-\Omega t)}\omega_{n'}Q_{nn'}b+c.c.]a_{n'}$ per optical mode that takes part in the Brillouin scattering process, when we consider variations in $\mathbb{E}$ due to an acoustic field, with $v_n$ being the group velocity of the mode, $\Gamma_n$ being the amplitude damping parameter, $\phi_n = k_n z-\omega_n t$ and $Q_{nn'}$ denotes the opto-mechanical perturbation overlap that characterises the interaction between distinct optical states mediated by the acoustic field. 
The equations of linear elasticity, which connect strain and mechanical displacement, underlie the dynamics on the acoustic front : $S = \nabla_s U$, $\rho\delta_t^2 U = \nabla.T+f$,
where $S$ is the linear strain tensor, $f$ is the body force density, $\nabla_s$  denotes the symmetric gradient operator, $U$ stands for the infinitesimal mechanical displacement vector, $\rho$ is the mass density and $T$ denotes the Cauchy stress. The equation for the acoustic mode is then given by: $(\delta_t + v_b\delta_z + \Gamma_b)b =\frac{1}{E_b}e^{-i(qz-\Omega t)}\Big[\sum_{nn'}e^{i(\phi_n-\phi_{n'})}a_na_{n'}^{*} \nonumber\langle\Phi\vert F_{nn'}\rangle+e^{-i(\phi_n-\phi_{n'})}a_n^{*}a_{n'}\langle\Phi\vert F_{n'n}\rangle\Big]$, where $\vert F\rangle = [0, f]^{T}$ is the driving term due to optical forces (fluctuations of optical intensity).
\subsection{Magnetic Flux as Mediator of Acousto-electric Coupling}

Inductance is the tendency of an electrical conductor to resist a change in the electric current that flows through it\,\cite{jones2013theory}. A magnetic field is produced around a conductor by an electric current flowing through it. The magnitude of the current determines the field strength, which follows any variations in current. Electromagnetic induction is the process by which a change in the magnetic field flowing through a circuit leads to the induction of an electromotive force in the conductors. The cylindrical coil is an idealized form of inductor having, on a cylindrical surface, a single-layer winding of constant axial pitch\,\cite{wheeler1958spherical}. If we have a core of a  rare-earth aluminosilicate glass, we see a diamagnetic magnetic response at the centre, which is altered based on geometric variations due to the acoustic mode\,\cite{so2019magnetic, majerova2022structure}. In a rare-earth aluminosilicate glass, the bonding characteristics of the $[\mathrm{SiO}_4]$ can be studied using NMR and IR spectroscopy\,\cite{kohli1992structural, kohli1993raman}. In systems like $\mathrm{CeO}_2-\mathrm{Al}_2\mathrm{O}_3-\mathrm{SiO}_2$ glasses, the aluminium is seen to have Al(4) and Al(6) coordination, using NMR studies\,\cite{lin1996structures}. In NMR studies, the Al(IV) lines show up as peaks at 54.7 ppm and 54.5 ppm for lanthanum and ytterbium glasses respectively\,\cite{marchi2005characterization}. The lanthanum-glasses possess a mixed silica-alumina network with dispersed silicate groups, while other rare-earth aluminosilicate glasses, the network is constituted primarily of silicate groups, with $\mathrm{Al}^{3+}$ as modifiers. A decrease in magnetic permeability was found in polycrystalline Ni-Zn ferrites (NZFOs) upon increasing $\mathrm{Al}^{3+}$ substitution in $\mathrm{Al}^{3+}$ substituted $\mathrm{Ni}_{0.7}\mathrm{Zn}_{0.3}\mathrm{Fe}_2\mathrm{O}_4$ nanoparticles\,\cite{birajdar2012permeability}. The silica in the system has a near-unity permeability which may be reduced by the presence of the $\mathrm{Al}^{3+}$. 
\begin{figure*}
    \centering    \includegraphics[scale=0.40]{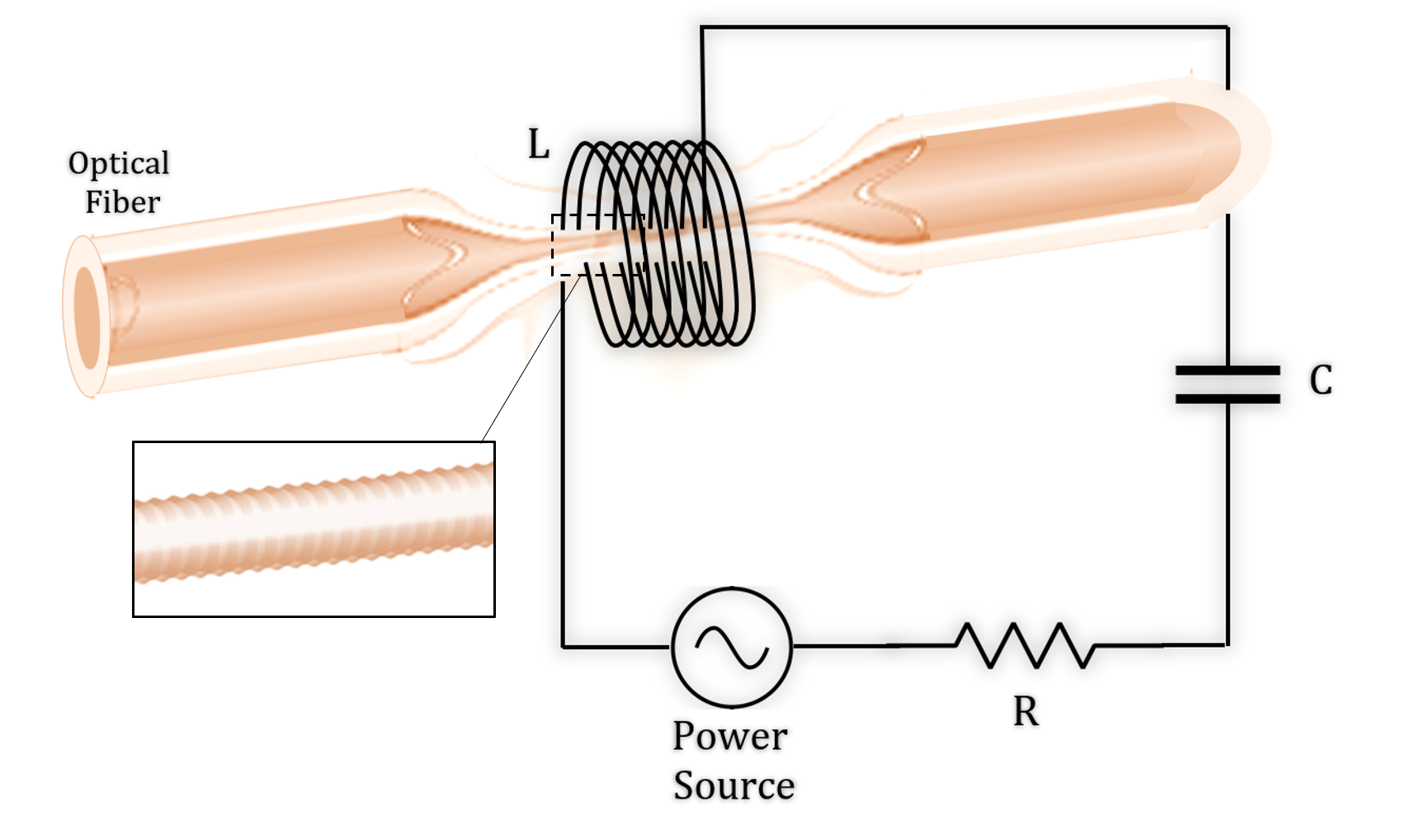}
    \caption{The scheme for quantum transduction using optoelectromechanical coupling of light in an optical fiber with a microwave signal through the acoustic mode generated in Brillouin scattering in a rare-earth aluminosilicate glass microwire. In this setup, the tapering of an optical fiber carrying the optical signal comprises of the microwire, which is taken as the core of a coil-inductor that is part of microwave resonator RLC circuit. The variation of the cross-sectional area temporally, due to the acoustic mode, leads to variations of magnetic flux and, as a result, the electrical response in the microwave resonator.}
    \label{fig:scheme}
\end{figure*}
\section{Results and Discussion}
The primary element in our scheme involves a coil-inductor with a sub-wavelength rare-earth aluminosilicate glass microwire core that facilitates generation of surface acoustic waves using Brillouin scattering. Rare-earth aluminosilicate glasses have high glass transformation temperatures and moderate thermal coefficients of expansion \cite{shelby1990rare}. With high concentrations of rare-earth ions, these glasses have excellent weathering tendency. Interestingly, the glass properties like refractive index, dilatometric softening, and glass transformation, temperatures have a linear variation with respect to the rare-earth ionic radius. It has been hypothesized that the field strength of these ions is the underlying basis for such variations. Since the field strength is related to the ionic radius via an inverse relation, we cannot separate the significance of the two. 
\\
\\
The schematic illustration of our  scheme is presented in Fig.\,\ref{fig:scheme} where rare-earth aluminosilicate glass microwire is used as core. This coil is part of the RLC microwave resonator circuit. The surface acoustic waves in the sub-wavelength optical fiber can be effective tools to mediate inductively between the incoming optical signals and a microwave resonator. If we assume that the acoustic wave is phase-matched with the beat between the first two optical modes such that $\Omega = \omega_2-\omega_1$ and $q = k_2 - k_1$, we can assume the driving force density $\textbf{F}$ to be of the form
\begin{equation}
    \textbf{F}(\textbf{r},t) = \textbf{f}(\textbf{r},t)a_1^*a_2 + c.c.
\end{equation}
If we were to consider the acoustic damping due to viscous forces $\eta$, we can write the acoustic mode equation as
\begin{equation}
    (\partial_z + \frac{1}{v_b}\partial_t + \alpha)b = -\frac{i\Omega a_1^*a_2}{\mathcal{P}_b}\mathcal{Q}_b
\end{equation}
where
\begin{equation}
    \alpha = \frac{\Omega^2}{\mathcal{P}_b}\Big[\int  d^2r\sum_{jkl}\vert\Phi\rangle_i^*\partial_j\eta_{ijkl}\partial_k\vert\Phi\rangle_l\Big],
\end{equation}
\begin{equation}
    \mathcal{Q}_b = \int d^2r (\vert\Phi(x,y)\rangle e^{i qz-i\Omega_n t})^*.\textbf{f}(r,t),
\end{equation}
with $\mathcal{Q}_b$ being the coupling parameter that is a work linear density, $\mathcal{P}_b$ being the mode’s energy transport velocity and $\alpha^{-1}$ is an effective dissipation length for the acoustic mode. We can similarly write the equations for the optical envelope functions as
\begin{equation}
    \partial_z a_1 + \frac{1}{v_1}\partial_t a_1 = - \frac{i\omega_1 a_2 b^*}{\mathcal{P}_1}\mathcal{Q}_1,
\end{equation}
\begin{equation}
    \partial_z a_2 + \frac{1}{v_2}\partial_t a_2 = - \frac{i\omega_2 a_1 b^*}{\mathcal{P}_2}\mathcal{Q}_2
\end{equation}
with 
\begin{multline}
    \mathcal{Q}_i = \int d^2r\Big[(\vert\Psi_{i}(x,y)\rangle e^{i k_i z-i\omega_i t})^*.\triangle \textbf{d}_i\\-\textbf{d}_i^*.\triangle \textbf{e}_i-\mu_0 \textbf{h}_i^*.\triangle \textbf{h}_i\Big]
\end{multline}
where $\textbf{d}_i$ is the induction-mode, $\triangle \textbf{d}_i$ is the perturbation of the same, $\textbf{h}_i$ is the magnetic field-mode with the $\triangle \textbf{h}_i$ being the perturbation thereof, and $\triangle \textbf{e}_i$ being the perturbation in the $i^\mathrm{th}$ electric field mode. The photoelastic effect, field perturbations due to varying continuity conditions and effective dynamic magnetic couplign leads to perturbations $\triangle \textbf{e}_i$, $\triangle \textbf{d}_i$ and $\triangle \textbf{h}_i$ respectively. Here, $\mathcal{P}_i$ gives the energy transport velocity of the $i^{\mathrm{th}}$ electromagnetic mode. Let us take the steady-state case and solve the equation for the acoustic mode by
means of its Green’s function $b(z) = -\frac{i\Omega \mathcal{Q}_b}{\mathcal{P}_b}\int_0^{\infty}dz'a_1(z-z')^*\times a_2(z-z')\exp(-\alpha z')$. Assuming that the optical powers change on a length scale that is much larger than $1/\alpha$, we can obtain the stimulated Brillouin scattering gain $G = \frac{\omega\Omega\mathcal{Q}_1\mathcal{Q}_b}{\mathcal{P}_1\mathcal{P}_2\mathcal{P}_b(\alpha - i\kappa)}$ with detuning parameter $\kappa$.
The generated acoustic modes lead to the cross-sectional area varying, leading to a change in the area of the inductor-core within the magnetic flux. In our system, we will be using a segment of the fiber taper as the dielectric core of a coil-inductor. The manner in which the geometric perturbations in the core can influence the flux can be seen by observing the changes in the electromotive force due to flux-variations: $\varepsilon = -\frac{d\Phi_B}{dt} = - \frac{d(\textbf{B}.\textbf{S})}{dt}$. For the finite coil in our system, we must consider the magnetic field in terms of the vector potential: $\textbf{B} = \nabla \times \textbf{A}$. Given the geometry, only the $\textbf{A}_\theta$ can be non-zero and $B_r = -\frac{\partial A_\theta}{\partial z}$, $B_z = \frac{1}{r}\frac{\partial (rA_\theta)}{\partial r}$, while for a solenoid made up of $n$ filaments per unit length, we have\,\cite{callaghan1960magnetic}
\begin{equation}
A_\theta = \frac{a\mu ni}{2\pi}\int_{\xi_-}^{\xi_+}d\xi\int_0^\pi \frac{\cos{\theta}d\theta}{\sqrt{\xi^2+r^2+a^2-2ar\cos{\theta}}}
\end{equation}
where $a$ is the radius of the coil, $\xi = y-l$ with $y$ is the axial coordinate while $l$ is the distance of the filament from the origin, $\xi_\pm = y\pm \frac{L}{2}$, $\mu$ is the permeability and $i$ is the current in each filament. We can now find the magnetic-field components 
\begin{equation}
B_r = -\frac{a\mu ni}{2\pi}\int_0^\pi\Big[\frac{\cos{\theta}d\theta}{\sqrt{\xi^2+r^2+a^2-2ar\cos{\theta}}}\Big]_{\xi_-}^{\xi_+}
\end{equation}
\begin{multline}
B_z = \frac{a\mu ni}{2\pi}\int_0^\pi \Big[\frac{\xi}{r^2 + a^2 - 2ar\cos{\theta}}\times\\ \frac{(a-r\cos{\theta})d\theta}{\sqrt{\xi^2+r^2+a^2-2ar\cos{\theta}}} \Big]_{\xi_-}^{\xi_+}
\end{multline}
For a surface acoustic wave travelling in the core such that the radius is a function of the axial distance $x(y,t) = A\sin{(2\pi y/\lambda-\omega t)}+r_i$ (where $A$, $\omega$ and $\lambda$ denote the amplitude, frequency and wavelength of the acoustic mode while $r_i$ denotes the undisturbed radius), we can write the area elements at the surface of the core as
\begin{equation}
da_z = \frac{da_r}{\cos{(\zeta)}}= \frac{r_i+A\sin(\zeta)}{\sqrt{1+\cos^2{\zeta}}} \lambda d\zeta
\end{equation}
where $\zeta=2\pi y/\lambda-\omega t$. Inductance is defined as $L = n\Phi_B/i$, and in this case, for $r_i \rightarrow 0$,
\begin{multline}
    L = -\frac{\mu n^2}{4}\Big[\frac{a^2r}{(\xi^2+a^2)^{3/2}}\Big]_{\xi_-}^{\xi_+}\int \frac{r_i+A\sin(\zeta)}{\sqrt{1+\cos^2{\zeta}}} \lambda \cos{(\zeta)}d\zeta\\ +\frac{\mu n^2}{2}\Big[\frac{\xi}{\sqrt{\xi^2+a^2}}\Big]_{\xi_-}^{\xi_+}\int\frac{r_i+A\sin(\zeta)}{\sqrt{1+\cos^2{\zeta}}} \lambda d\zeta
\end{multline}
If we consider $\xi>>a$ and $p^2 = 2-\sin^2{\zeta}$, with the integration over $p$ with limits $1\rightarrow \sqrt{2}$, along with considering $r_i = 1\,\mu$m and $A = 0.05\,\mu$m, we have $L=5.1 \ times 10^{-9}\mu$ for velocity of acoustic mode $v = 5727\,m/s$ and frequency $f = 11.476\,\mathrm{GHz}$, as is seen to be the case for a microwire ($r_i = 1\,\mu$m, $l = 8\,$cm) at incident light with wavelength 1534 nm (195.57 THz) in the case of lanthano-aluminosilicate\,\cite{dragic2014brillouin}, while $n=250$. 
For a resonator circuit with $C = 47\,pF$ and standard inductive element $L = 1\,$mH, we then have the frequency of 
\begin{equation}
    f_r = \frac{1}{2\pi\sqrt{LC}} = 325.08\,\mathrm{MHz}
\end{equation}
%
The primary hurdle when it comes to efficiency is the emergence of disparate acoustic waves besides the surface acoustic waves we require. The primary factors that lead to losses include the photon-to-acoustic mode conversion efficiency and the losses in the microwave resonator circuit. Beginning with the theory of Brillouin scattering, the Brillouin gain $g_{B}$ can be expressed as\,\cite{ippen1972stimulated} 
\begin{equation}
    g_B = \frac{2\pi n_{\mathrm{eff}}^7P_{12}^2}{c\rho\lambda^2v_s\triangle\nu_B}
\end{equation}
where $P_{12}$ is the elasto-optic constant, $\rho$ is the mean density of the material, $n_{\mathrm{eff}}$ is the effective refractive index of the microwire, $v_s$ is the sound velocity, $c$ is the speed of light in vacuum, $\nu_B$ and $\triangle \nu_B$ are the frequency shift of the surface acoustic wave and FWHM width of the spectrum respectively, while $\lambda$ is the optical wavelength. Rare-earth aluminosilicates have an effective refractive index of $n_{\mathrm{eff}} = 1.65$\,\cite{shelby1990rare}. The Pockel's constant is found to be $P_{12} = - 0.027\pm 0.012$\,\cite{dragic2013pockels}. Aluminosilicate glass has a mean density of $\rho = 3.48–4.19\pm 0.02\, \mathrm{g\, cm}^{-3}$\,\cite{marchi2005characterization}. For Yb-doped aluminosilicate glass, we have $\nu_B = 11$ GHz and $\triangle \nu_B = 287 - 1550$ MhZ\,\cite{dragic2013brillouin}. 
\\
\\
There have been studies on the variation of compressional wave velocity and shear wave velocity of four types of aluminosilicate glasses as a function of pressure, and the velocity ranged from $\sim 5.5 - 8$ km/s for compressional wave velocity, ranged $\sim3.2 - 4.2$ km/s for shear wave velocity and ranged $\sim4 - 6.2$ km/s for bulk sound velocity for pressure varying over $0-8$ GPa\,\cite{aoki2020effects}. Using these values, we obtain the Brillouin gain of $g_B = 1.0727 \times 10^{-11}$ m/W. From literature, a Yb-doped aluminosilicate fiber with 5 mol\% of alumina and 0.2 mol\% of Ytterbia has been seen to have a Brillouin Gain coefficient of $1.1 \times 10^{-11}$ m/W, while phosphosilicate fiber having 20 mol\% of $\mathrm{P}_2\mathrm{O}_5$ and 0.6 mol\% of Ytterbia, the BGC is seen to be around $0.7\times 10^{-11}$ m/W\,\cite{dragic2013brillouin}. The critical power is given by \cite{pant2011chip}:
\begin{equation}
    P_{\mathrm{cr}} = 21 \frac{A_{\mathrm{eff}}}{\kappa g_BL_{\mathrm{eff}}}
\end{equation}
where $\kappa$ is the parameter that accounts for variations in polarisation. For our purposes, we take $\kappa$ to be unity. If we take a microwire with $1\,\mu$m diameter over a length of 8 cm, $P_{\mathrm{cr}} = 19.2194$ W. We can describe the spontaneous emission noise by adding a spontaneous emission term to the
equations that describe how the pump and Stokes waves evolve\,\cite{ferreira2011optical}. The amplified spontaneous scattered
power per unit frequency is given by
\begin{multline}
    P(\nu,z)= \frac{\hbar \nu}{A_e}\frac{\Gamma^2}{\delta^2+\Gamma^2}g_B(N+1)G(\nu,z) \int_{z}^{L}\frac{P_p(z')}{G(\nu,z')}dz'
\end{multline}
\begin{equation}
    G(\nu,z) = \exp{\Big[\int_{z}^{L}\Big(\frac{P_p(z)}{A_e}\frac{\Gamma^2}{\delta^2+\Gamma^2}g_B-\alpha\Big) dz]\Big]}
\end{equation}
where $A_e$ is the effective core area, $\Gamma^{-1}$ is the acoustic phonon lifetime, $\delta$ parametrizes any detuning of the signal from the Stimulated Brillouin Scattering gain line center, $P_p$ is the pump wave power and $\alpha$ is the absorption coefficient. $N \approx \frac{kT}{h\nu_B}$ denotes the number of acoustic phonons in thermal equilibrium, with $k$ being the Boltzmann constant and $T$ being the absolute temperature. In our scheme, if we take $\nu_B = 11.476\,\mathrm{GHz}$ at $T=298$ K, $N \approx 565$ phonons. Fiber-ring lasers can have pulse duration of 4-30 ps\,\cite{yao2022high}. Specific kinds of systems such as whispering-mode stimulated Brillouin scattering has shown a field distribution for the photonic and phononic modes, with a maximum photon-to-phonon conversion of $\sim 45$\%. If the phonon-to-microwave conversion is optimum, we can regard the overall efficiency to be comparable to the existing schemes for quantum transduction. In this setup, the noise-power significantly depends on the magnitude of the signal in the gain-saturation regime. It is particularly high when the signal has low power. This imposes certain constraints on such a Brillouin-scattering based system.

 \section{Conclusion}
We have proposed a scheme for realisation of optoelectromechanic quantum transduction using rare-earth aluminosilicate glass microwires as the dielectric core of a coil-inductor.  Our scheme makes use of  Brillouin interactions in a fiber mode that is allowed to be passed through a fiber taper. We have shown how we can convert a light at 195.57 THz to a microwave signal at 325.08 MHz. We see that the Brillouin gain coefficient obtained for rare-earth aluminosolicates of $1.0727\times 10^{-11}\,$m/W can give a critical power of $P_{cr}=19.2194$ W. Given the ease of access and manoeuvrability of microwires in  experimental setups,  it can be a practically relevant scheme for quantum transduction and can be further engineered for desired transduction. For a whispering SBS mode, we can attain upto $45$\% conversion efficiency using  optoelectromechanical transduction scheme presented in this paper.\\

\vskip 0.2in

{\bf Acknowledgement :}
MGM and CMC acknowledge the support from the Office of Principal Scientific Advisor to Government of India, Project No. Prn.SA/QSim/2020\\

{\bf Data Availability Statement :}
Data sharing not applicable to this article as no datasets were generated or analysed during the current study.
\bibliography{apssamp}

\end{document}